\documentclass{aac}

\usepackage[numbers]{natbib}
\usepackage{xcolor}
\usepackage{rotating}
\usepackage{graphicx}
\usepackage{siunitx}
\usepackage{bm}

\begin{document}

% Title portion
\title{Plasma Wakefield Accelerators with Ion Motion and the E-314 Experiment at FACET-II}

\author[aff1]{Claire Hansel\corref{cor1}}
\author[aff1,aff2,aff3]{Monika Yadav}
\author[aff1]{Pratik Manwani}
\author[aff1]{Weiming An}
\author[aff1]{Warren Mori}
\author[aff1]{James Rosenzweig}

\affil[aff1]{Department of Physics and Astronomy, University of California Los Angeles, California 90095, USA}
\affil[aff2]{Cockcroft Institute, Warrington WA4 4AD, UK}
\affil[aff3]{Department of Physics, University of Liverpool, Liverpool L69 3BX, UK}
\corresp[cor1]{clairehansel3@gmail.com}

\maketitle

\begin{abstract}
A future plasma based linear collider has the potential to reach unprecedented energies and transform our understanding of high energy physics. The extremely dense beams in such a device would cause the plasma ions to fall toward the axis. For more mild ion motion, this introduces a nonlinear perturbation to the focusing fields inside of the bubble. However, for extreme ion motion, the ion distribution collapses to a quasi-equilibrium characterized by a thin filament of extreme density on the axis which generates strong, nonlinear focusing fields. These fields can provoke unacceptable emittance growth that can be reduced through careful beam matching. In this paper, we discuss the rich physics of ion motion, give a brief overview of plans for the E-314 experiment at FACET-II which will experimentally demonstrate ion motion in plasma accelerators, and present results of particle-in-cell simulations of ion motion relevant to the E-314 experiment. 
\end{abstract}

\section{INTRODUCTION}

Plasma wakefield accelerators (PWFAs) have demonstrated the ability to accelerate particles with unprecedented acceleration gradients, far in excess of those achievable with conventional radio frequency cavities \cite{nature}. Much research on PWFAs has focused on the blowout regime \cite{jamieblowout} in which the electron beam density is much greater than the plasma density. In this regime, the plasma electrons are repelled away from the axis by the beam and crash back toward the axis after the beam has left, creating a bubble free of plasma electrons which trails the beam. Due to their much larger mass, plasma ions are typically assumed to be stationary inside the bubble and thus generate linear focusing fields which are desirable for preserving beam emittance. In some proposals, such as the plasma afterburner \cite{afterburner1,afterburner2}, this stationary ion assumption breaks down because the beam is dense enough and long enough to cause the ions to fall toward the beam \cite{jamieionmotioncalc}. In this paper we discuss the physics of ion motion, the planned E-314 experiment at FACET-II, and particle-in-cell (PIC) simulations of ion motion.

\section{PHYSICS OF ION MOTION}

The degree of ion motion in a plasma accelerator can be quantified by a dimensionless parameter known as the \textit{phase advance} \cite{jamieionmotioncalc}:

\begin{equation} \label{eq:phaseadvance}
\Delta \phi = 2 \pi  \sigma_z \sqrt{\frac{r_e Z_i n_{b,0} m_e}{m_i}},
\end{equation}

\noindent where $r_e$ is the classical electron radius, $Z_i$ is the ion charge state, $m_i$ is the ion mass, $n_{b,0}$ is the number density of electrons at the center of the beam, and $\sigma_z$ is the rms length of the beam. $\Delta \phi$ is derived by considering a beam with a cylindrically symmetric bigaussian transverse distribution of spot size $\sigma$ and a flat top longitudinal distribution of length $\sqrt{2 \pi} \sigma_z$. Ions begin approximately at rest, and the strongest attraction toward the axis is experienced by ions which have initial radius near the center of the beam ($r_0 \ll \sigma$) where the beam density is approximately constant ($n_b \simeq n_{b,0}$). Thus the trajectory of the ions is given by $r(\xi) = r_0 \cos(k_i \xi)$ where $\xi = ct - z$ and $k_i = \sqrt{2 \pi r_e Z_i n_{b,0} m_e / m_i}$. The position of the ions at the end of the beam is $r_{\text{end}} = r_0 \cos(\Delta \phi)$. The degree of ion motion can be divided into three regimes. For $\Delta \phi \ll \pi / 2$, ion motion is negligible, and the stationary ion approximation applies. For $\Delta \phi \lesssim \pi / 2$, ion motion is laminar, wavebreaking does not occur, and ion motion is relatively gentle. This regime was discussed in \cite{benedetti1} where expressions for the transverse wakefield and beam emittance growth were derived and emittance preservation via bunch shaping was discussed. Finally, for $\Delta \phi \gtrsim \pi / 2$, wavebreaking occurs and ion motion is no longer laminar. Ions pass through the axis and the ion phase space filaments on smaller and smaller scales as the ions collisionlessly relax towards a coarse grained equilibrium. This equilibrium is characterized by thin filament of extreme density on axis, and gives rise to strong, nonlinear focusing fields. The nonlinearity of these focusing fields causes filamentation of the beam's phase space which increases the beam's emittance. The beam and ion distributions do not evolve independently but instead evolve together as part of a coupled beam-ion system, eventually reaching an equilibrium. It was shown in \cite{an} that emittance growth can be reduced by focusing the beam to a spot size smaller than the matched spot size in the ion motion-free case. 

\section{E-314 EXPERIMENT}

E-314 is a planned experiment at the FACET-II facility at SLAC National Accelerator Laboratory which will experimentally demonstrate extreme ion motion. The preliminary experimental parameters are shown in Table \ref{tab:parameters}. In order to drive ion motion, a long, high charge, tightly focused beam will be used. For simplicity, a single beam will be used rather than two.

\begin{table}[h]
\begin{tabular}{|c|c|c|}
\hline
Parameter & Value & Unit \\
\hline
\multicolumn{3}{|c|}{Beam} \\
\hline
$E$ & $10$ & $\si{GeV}$  \\
$Q$ & $2.89$  & $\si{nC}$ \\
$\sigma_x,\sigma_y$ & $328$ & $\si{nm}$ \\
$\sigma_z$ & $40$ & $\si{\mu m}$ \\
$\epsilon_{n,x},\epsilon_{n,y}$ & $2$ & $\si{\mu m }$ \\
$\Delta \phi$ & $5$ & rad \\
\hline
\multicolumn{3}{|c|}{Plasma} \\
\hline
$n_0$ & $10^{18}$ & $\si{cm^{-3}}$ \\
Ion species & H$^{+}$ & \\
\hline
\end{tabular}
\caption{Preliminary parameters for the E-314 experiment.}
\label{tab:parameters}
\end{table}

Much of the experimental infrastructure for E-314 will be shared with other experiments at FACET-II. However, in order to achieve the demanding requirements for beam spot size, a triplet of strong ($>\SI{700}{T/m}$) permanent magnet quadrupoles \cite{magnet} will be added after the FACET-II final focus. A mirrored collimating triplet will be added downstream of the interaction point in order to propagate the beam to diagnostics including a beam imaging system and a momentum spectrometer. Another possibility is to use an underdense passive plasma lens \cite{lens} to achieve the small spot sizes required.

The oscillation of the beam electrons under the influence of the extremely strong nonlinear focusing fields produced by the ion column will emit a unique betatron radiation signature which will be measured by a novel spectrometer under development at UCLA \cite{monika}. This spectrometer will measure beam radiation in a single shot, non-destructive fashion for E-314 as well as other experiments at FACET-II. Evidence of ion motion will also be measured in the ions themselves: a compact ion dipole spectrometer will be used to analyze the kinetic energies of ions.

The particular experimental goals of E-314 are to (a) demonstrate the formation of a beam-ion equilibrium in agreement with PIC simulations and theory and (b) demonstrate emittance conservation through matching of the beam to the nonlinear focusing fields. These are ambitious goals which will represent significant progress towards a future plasma based linear collider.

\section{PIC SIMULATIONS}

We performed simulations of extreme ion motion for the preliminary parameters of the E-314 experiment shown in Table \ref{tab:parameters} using the quasi-static particle-in-cell code QuickPIC \cite{quickpic1, quickpic2}.

\begin{figure}[h]
    \centering
    \includegraphics[width=0.25\textwidth]{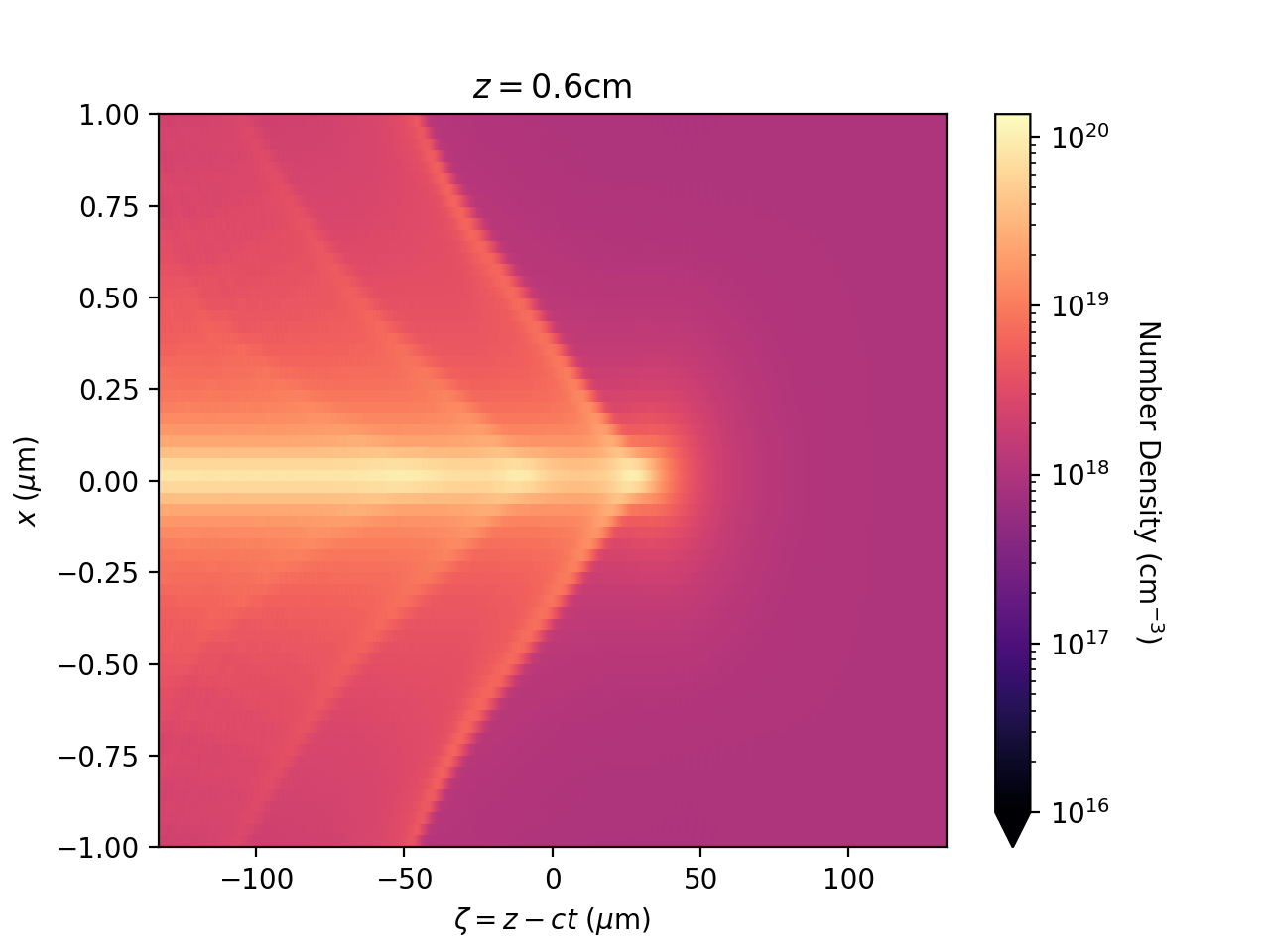}
    \includegraphics[width=0.25\textwidth]{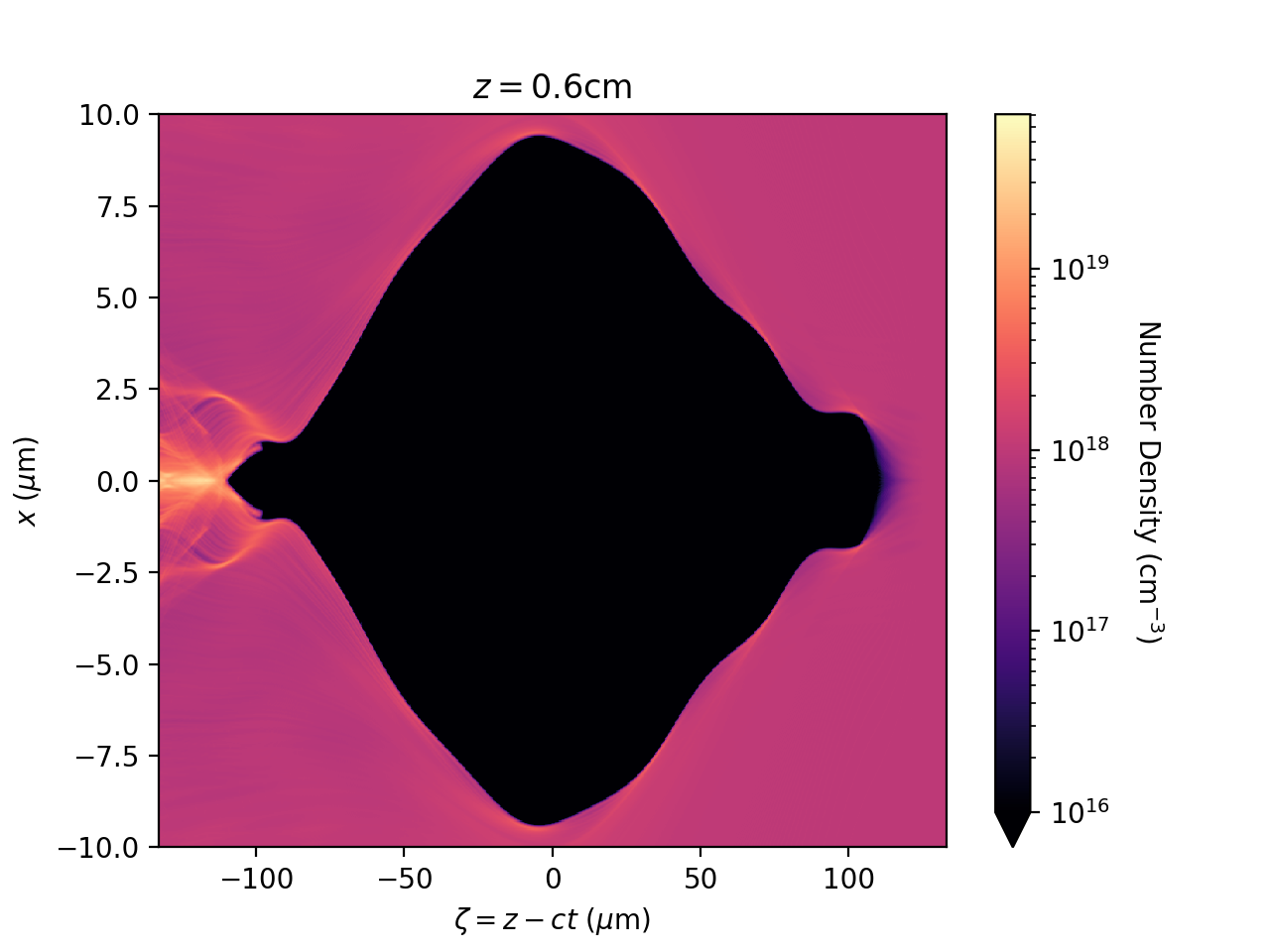}
    \includegraphics[width=0.25\textwidth]{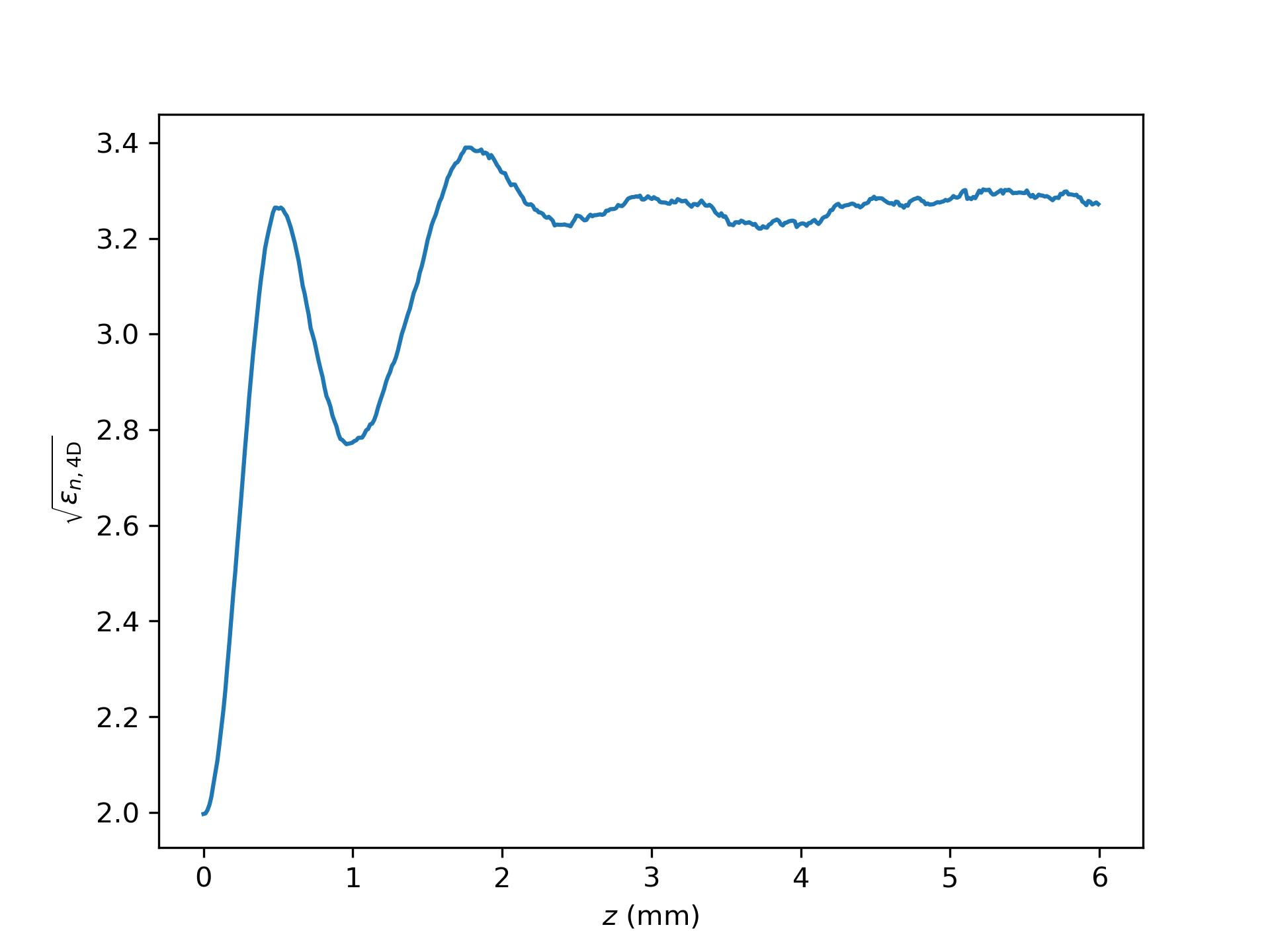}
    \includegraphics[width=0.25\textwidth]{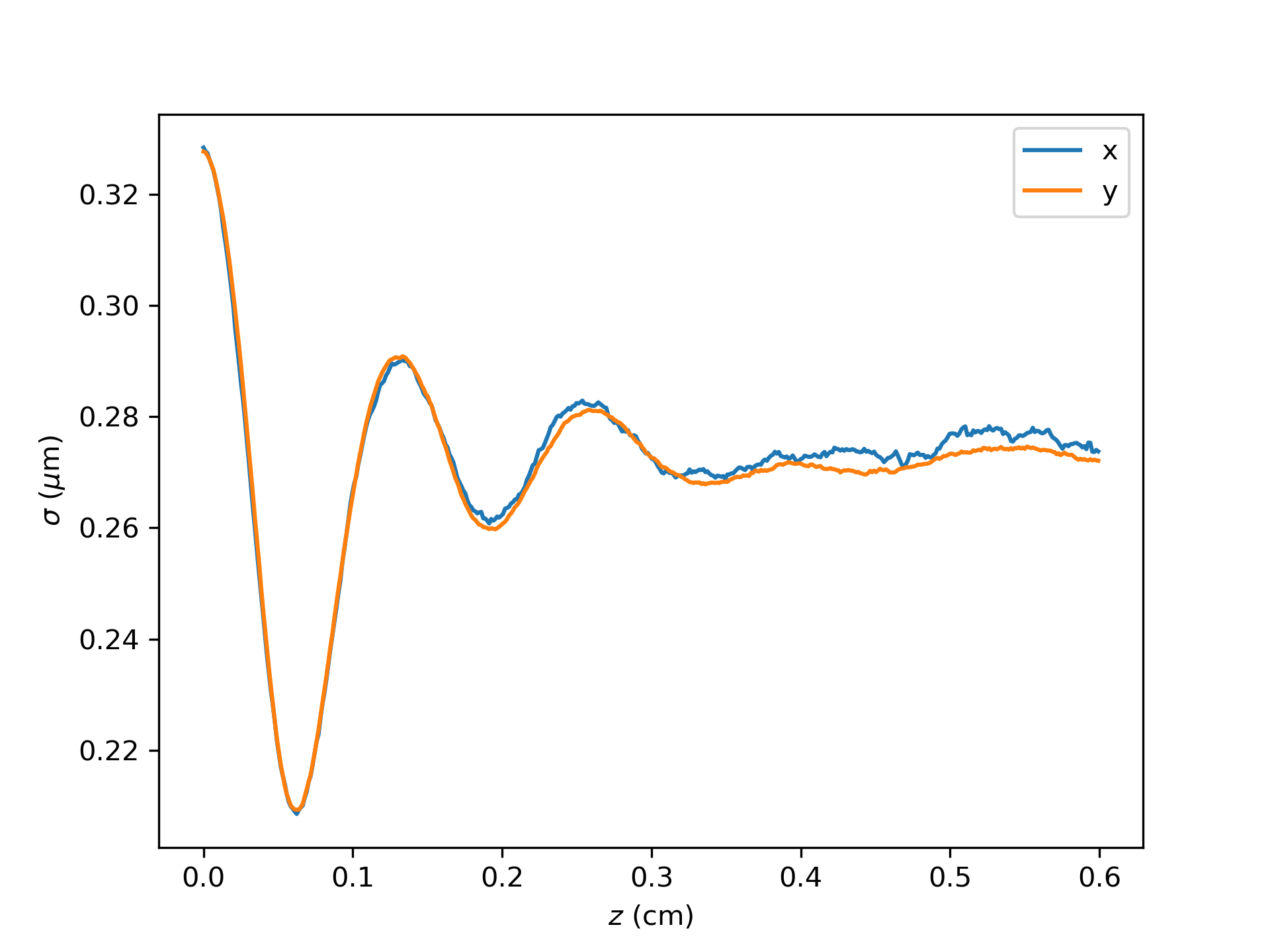}
    \centering
    \caption{\textbf{Far Left:} Plasma ion number density  slice at the end of the plasma. \textbf{Middle Left:} Plasma electron number density  slice at the end of the plasma. \textbf{Middle Right:} Beam emittance evolution over the length of the plasma. \textbf{Far Right:} Beam spot size evolution over the length of the plasma.}
    \label{fig:e314other}
\end{figure}

\begin{figure}[h]
    \centering
    \includegraphics[width=0.25\textwidth]{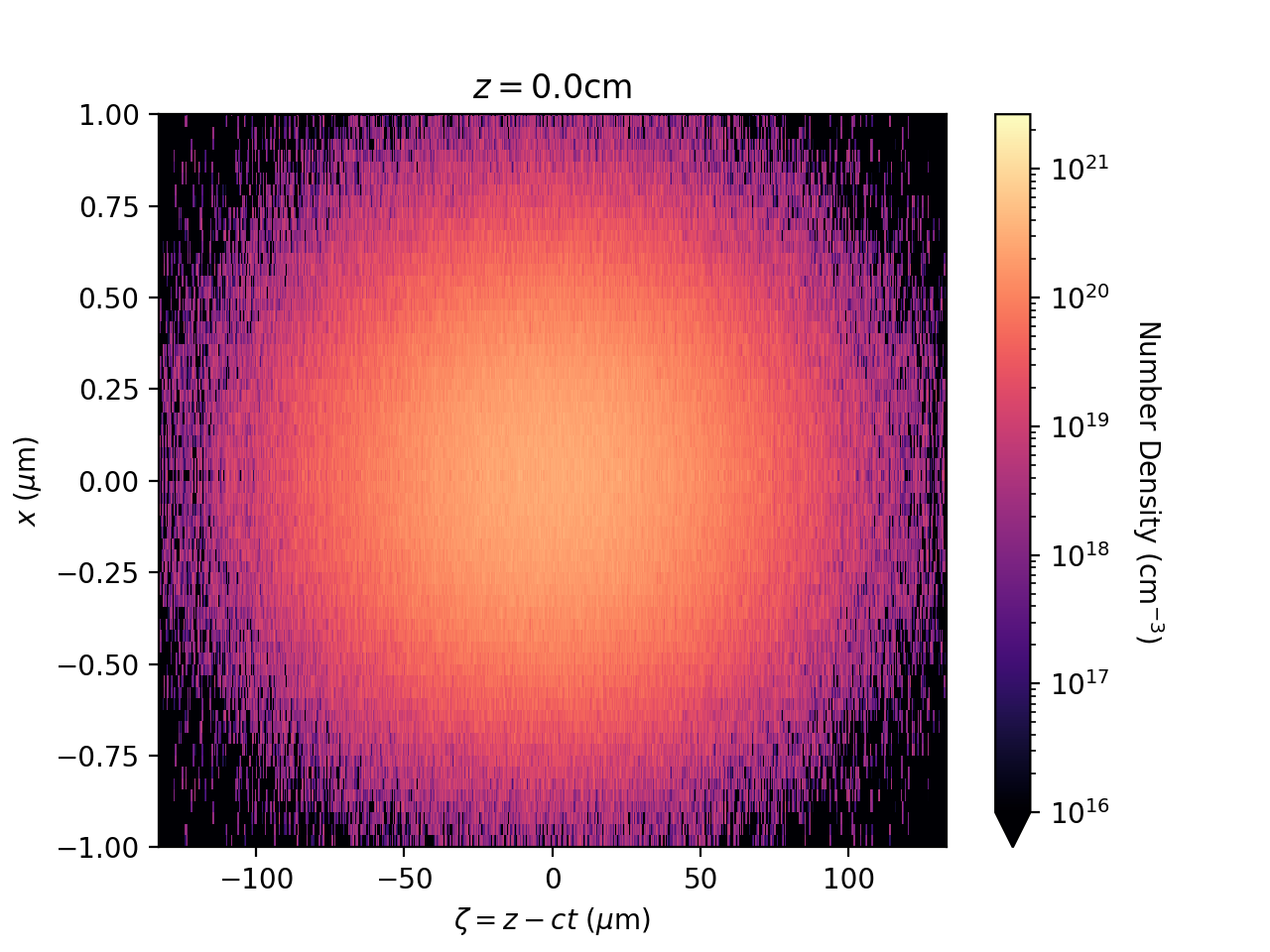}
    \includegraphics[width=0.25\textwidth]{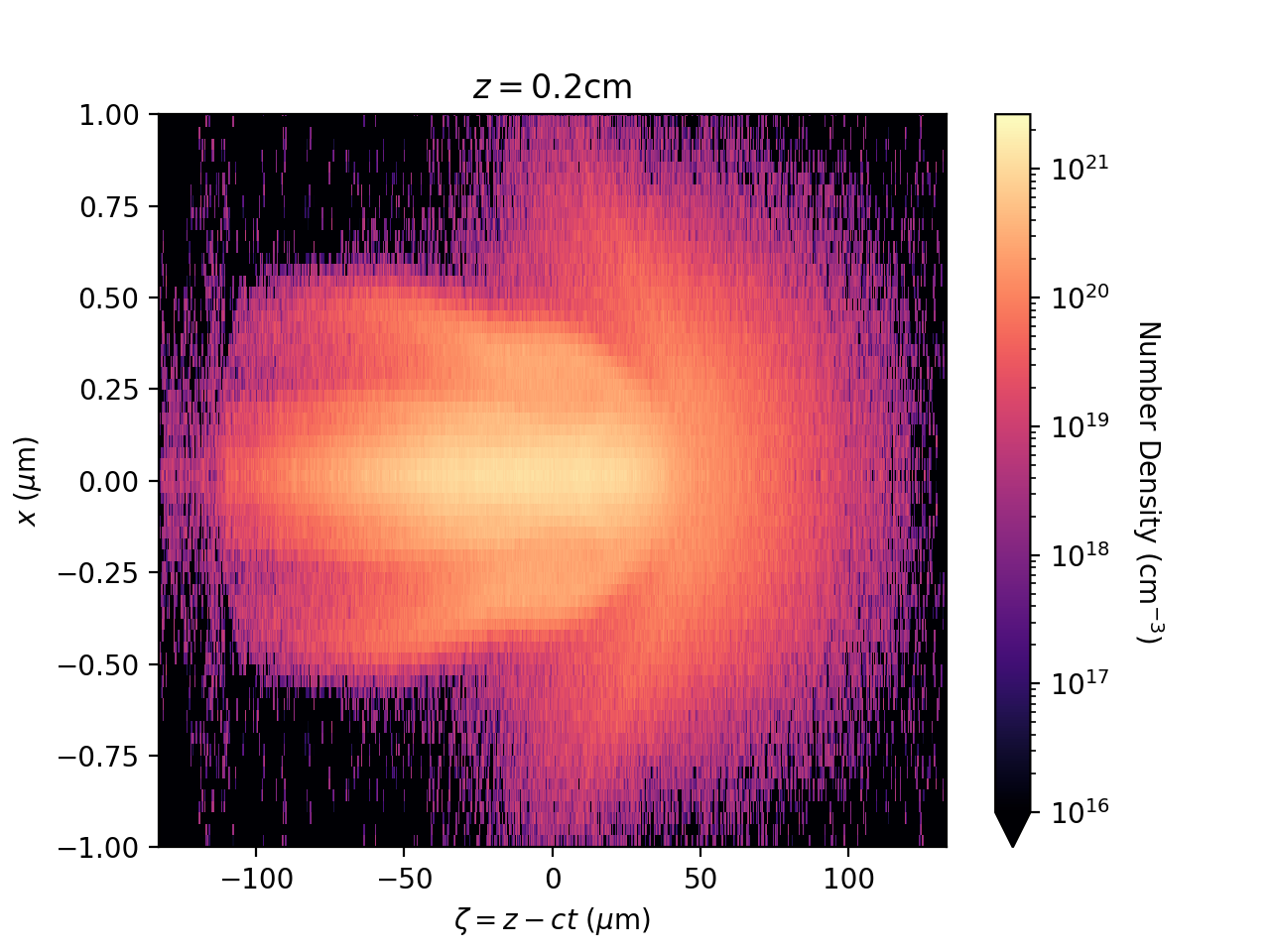}
    \includegraphics[width=0.25\textwidth]{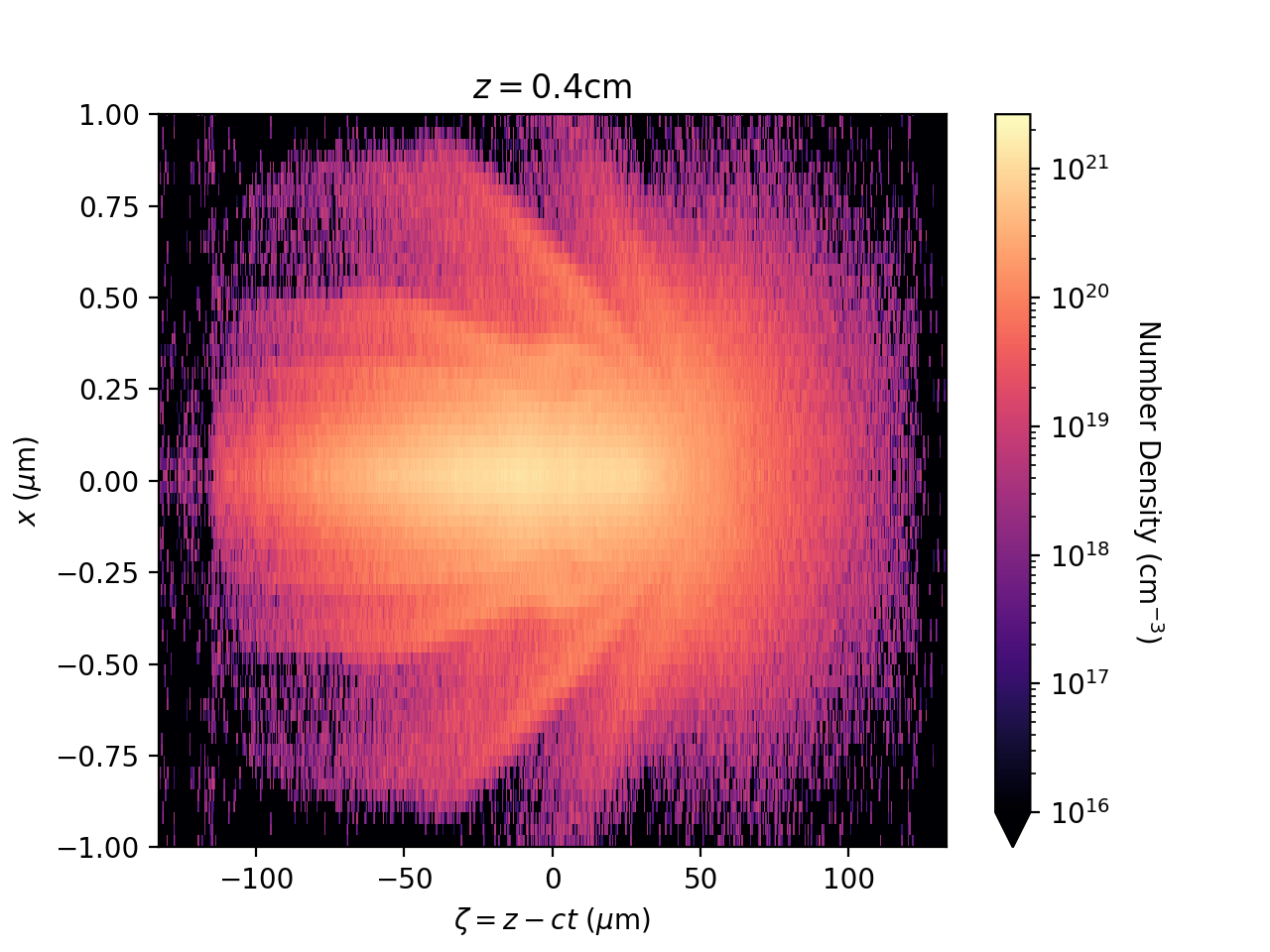}
    \includegraphics[width=0.25\textwidth]{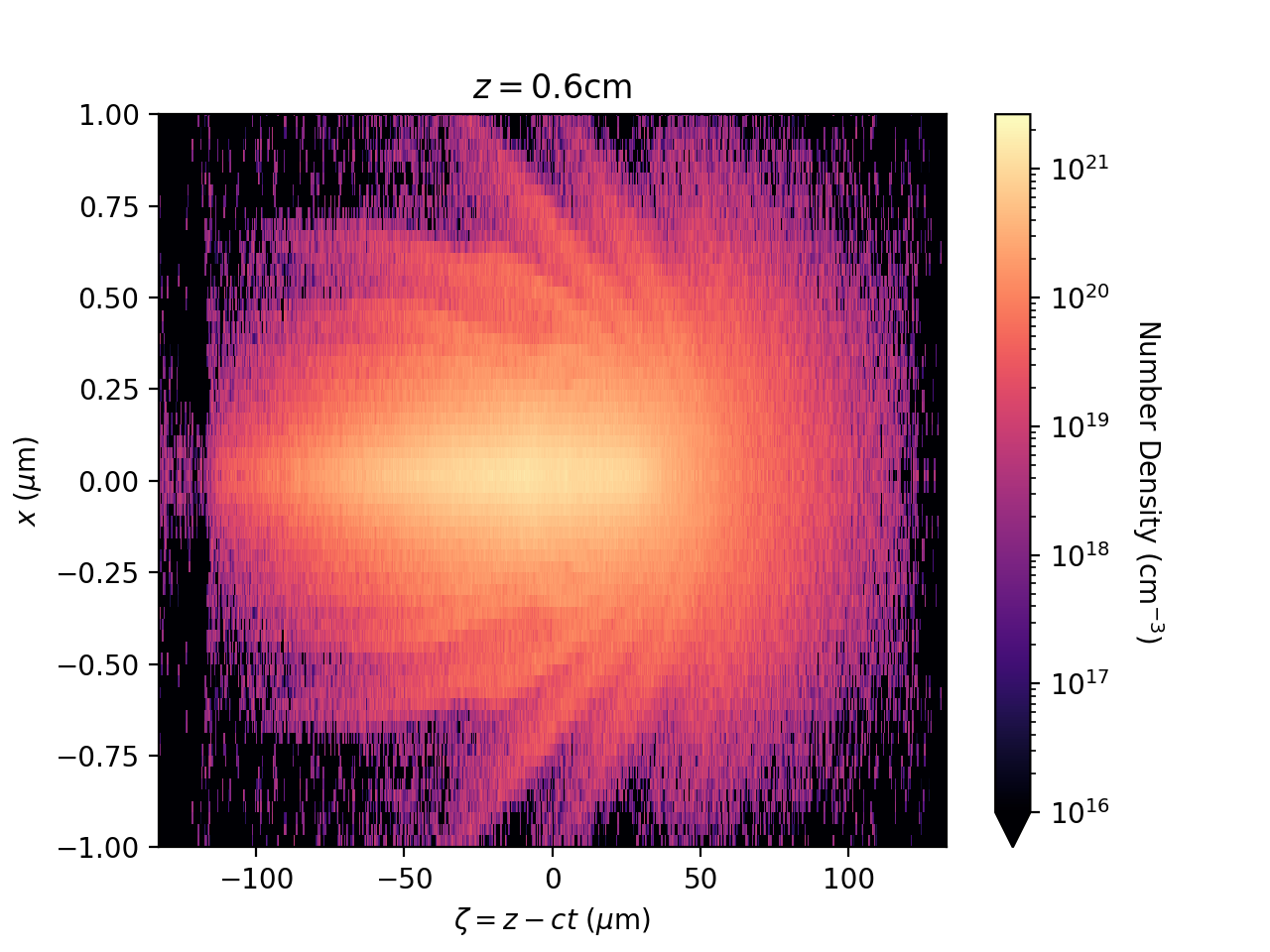}
    \centering
    \caption{Beam electron number density slice at $z = \SI{0}{mm}$ \textbf{(Far Left)}, $z = \SI{2}{mm}$ \textbf{(Middle Left)}, $z = \SI{4}{mm}$ \textbf{(Middle Right)}, and $z = \SI{6}{mm}$ \textbf{(Far Right)}.}
    \label{fig:e314beam}
\end{figure}

First, the beam evolution was simulated over a distance of $\SI{6}{mm}$. A transverse cell size of $\SI{31}{nm}$, a longitudinal cell size of $\SI{259}{nm}$, and a step size for beam evolution of $\SI{13}{\mu m}$ were used. The far left plot in Fig.~\ref{fig:e314other} shows the plasma ion density at the end of the plasma. Extreme ion motion can be clearly seen in the plasma ion response: the originally uniform ions collapse into a thin, dense ion column. Wakes extending away from the ion column are visible after wave breaking. Plasma electron blowout can be clearly seen in the middle left plot of Fig.~\ref{fig:e314other} which shows the plasma electron number density at the end of the plasma. Interestingly, the bubble is not completely round as is the case in the absence of ion motion. The evolution of the beam emittance and spot size are shown on the middle right and far right plots respectively of Fig.~\ref{fig:e314other}. The nonlinear focusing fields of the collapsed ion distribution cause mixing of the beam's phase space which leads to emittance growth. The emittance approaches a final saturated value as the beam reaches a coarse grained equilibrium. The evolution of the beam spot size shows a damped oscillation which tends towards a value smaller than the original spot size of $\SI{328}{nm}$. The original spot size is chosen because it is matched to the linear focusing field that would exist in the absence of ion motion. The collapse of the ions into a column produces stronger focusing fields which focus the beam tighter than those linear fields. Figure \ref{fig:e314beam} shows the evolution of the beam electron number density as the originally Gaussian beam approaches a coarse grained equilibrium. The head of the beam is constantly disturbing new ions which begin uniformly distributed and then collapse over the length of the beam. Because of this, the focusing field at the head of the beam is the same linear focusing field that, in the absence of ion motion, would exist over the entirety of the beam. Because the beam spot size is matched to this field, the head of the beam does not evolve over the length of the plasma, which can clearly be seen in the beam electron number density plots. By the middle and tail of the beam, the ions have had time to collapse and thus produce strong nonlinear fields which focus the beam tighter and cause mixing of the beam's phase space which can be clearly seen.

It is possible to ignore the plasma electron response when simulating ion motion provided the beam is entirely in the bubble. This approximation was employed by \cite{an} in the context of a drive and witness pair where the drive beam served to eject the plasma electrons but did not drive ion motion. In our single beam case, this approximation accurately represents all but the very front of the beam, which is analogous to to the drive beam. Simulations using this approximation provide important insights into the underlying physics of ion collapse as a smaller simulation window can be used which allows the ion column to be resolved much better. 

\begin{figure}[h]
    \centering
    \includegraphics[width=0.333\textwidth]{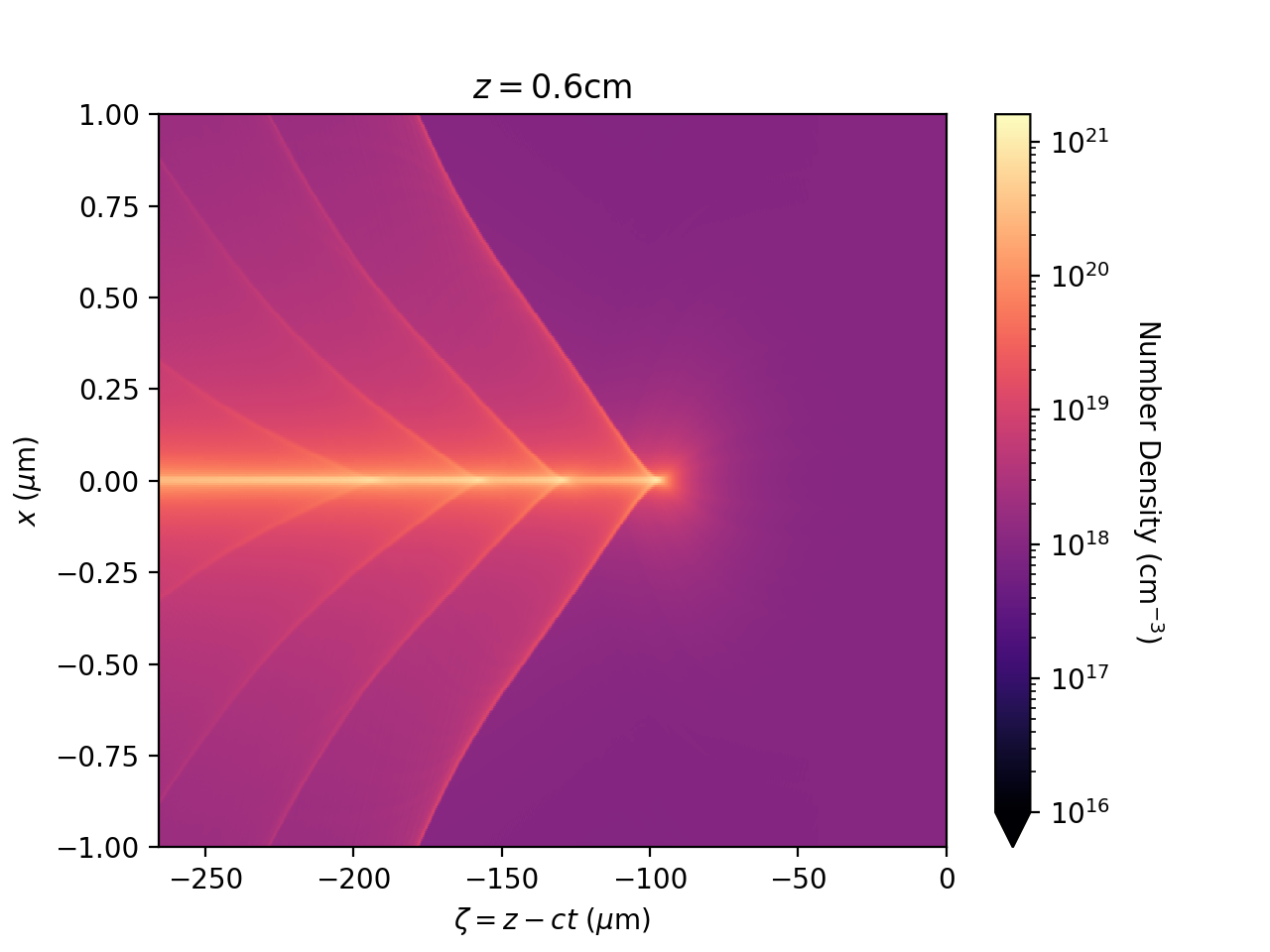}
    \includegraphics[width=0.333\textwidth]{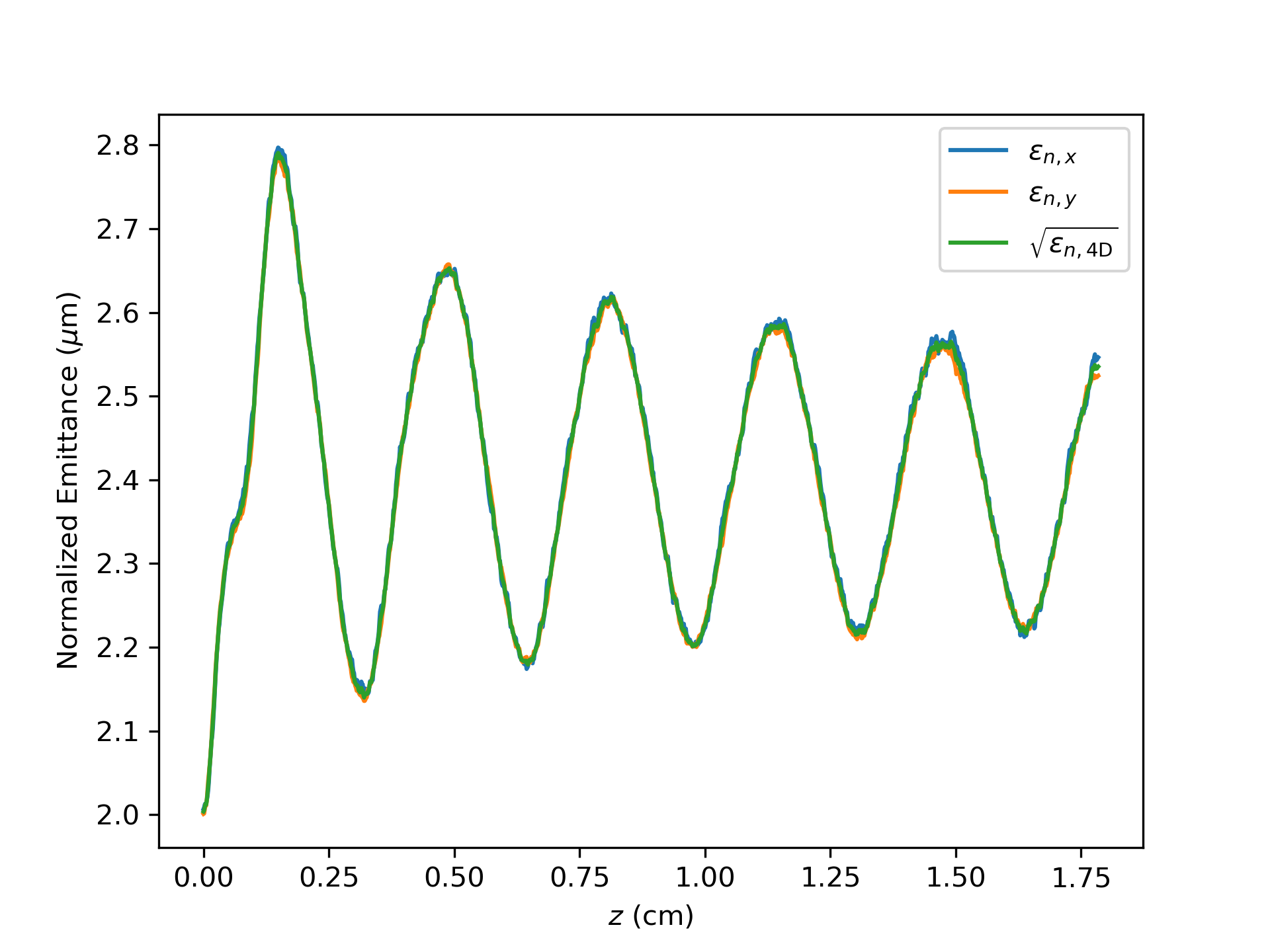}
    \includegraphics[width=0.333\textwidth]{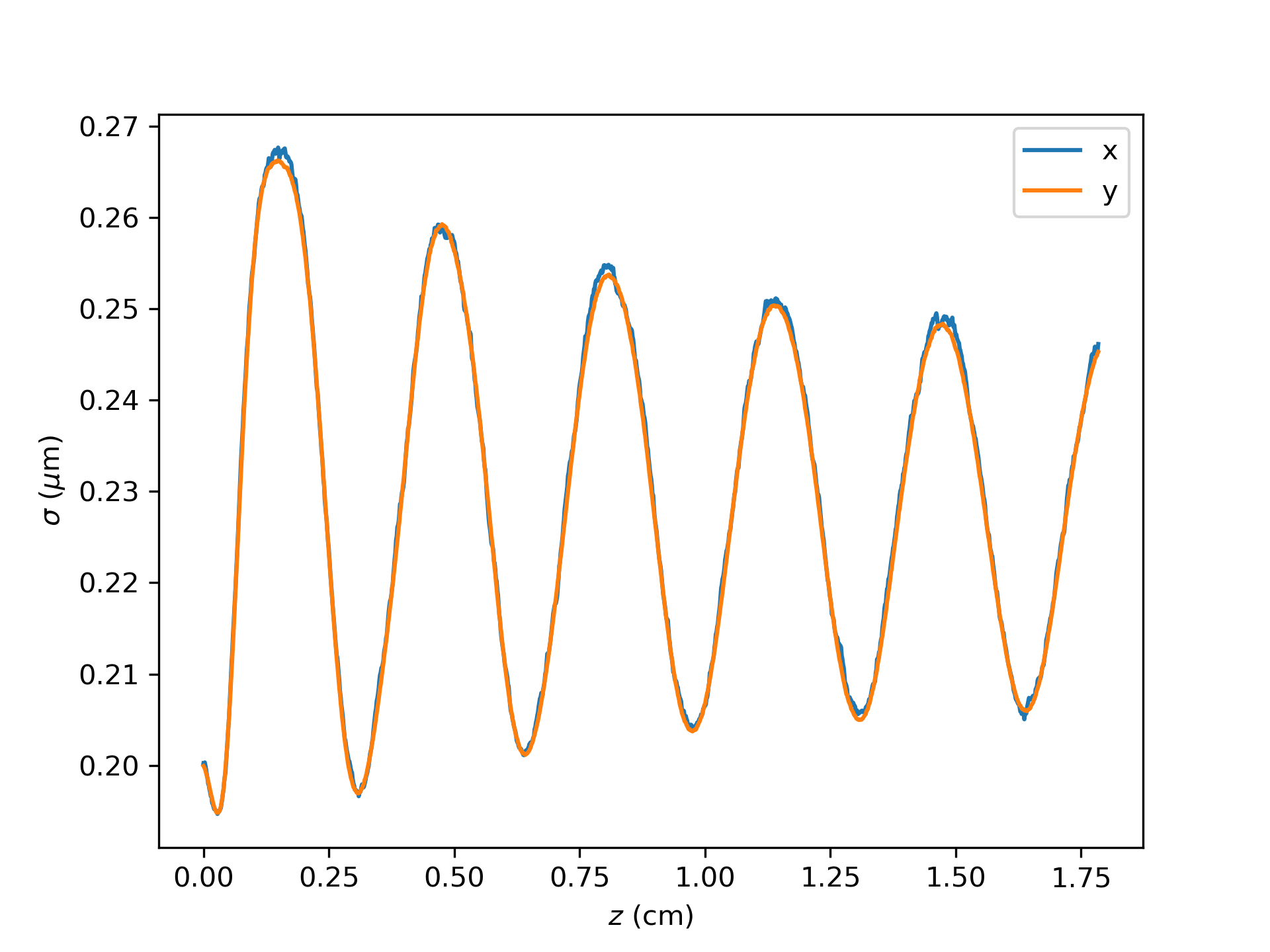}
    \centering
    \caption{\textbf{Left:} Plasma ion number density slice at the end of the plasma. \textbf{Middle:} Beam emittance evolution over the length of the plasma. \textbf{Right:} Beam spot size evolution over the length of the plasma.}
    \label{fig:other}
\end{figure}

\begin{figure}[h]
    \centering
    \includegraphics[width=0.25\textwidth]{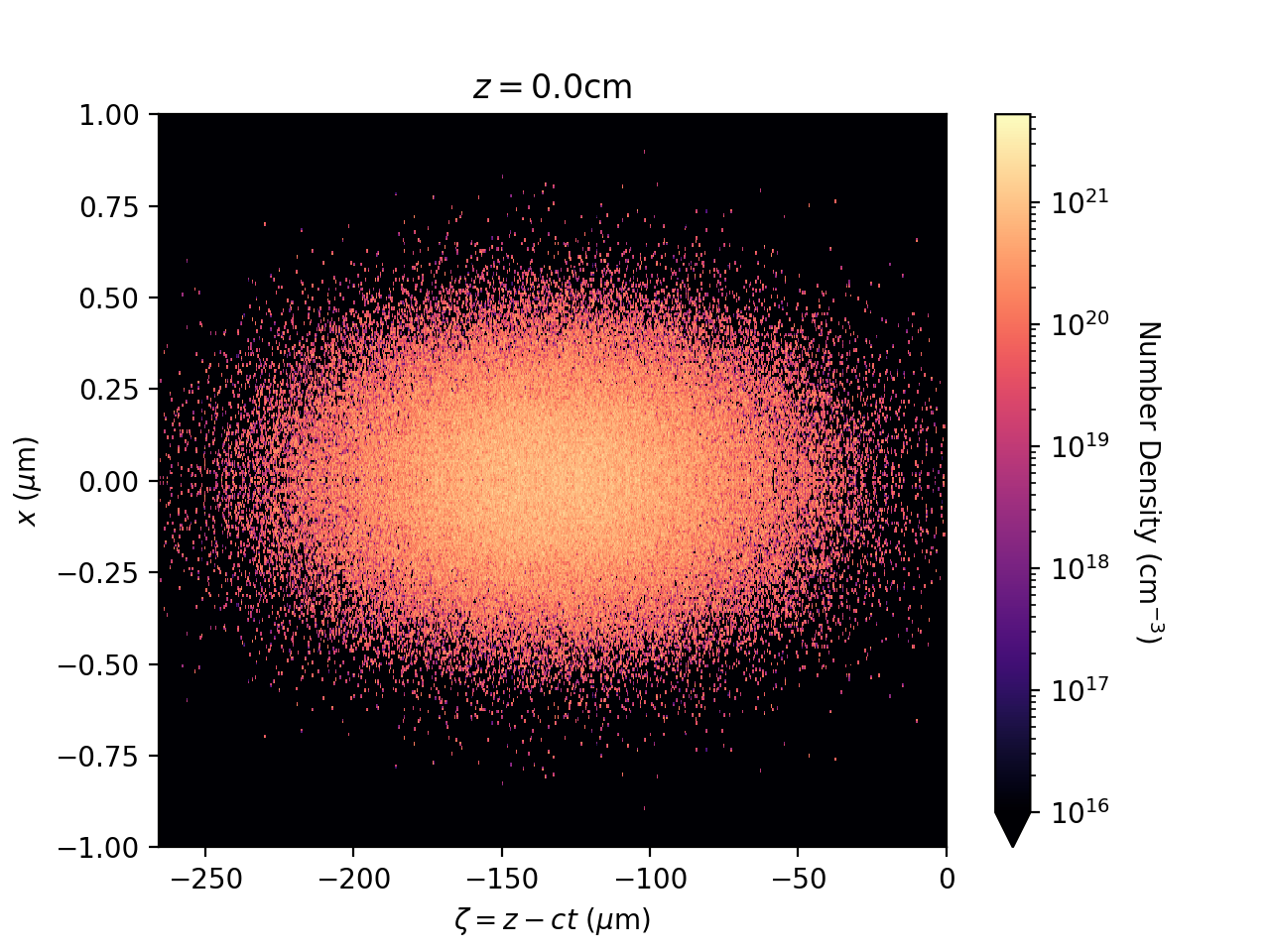}
    \includegraphics[width=0.25\textwidth]{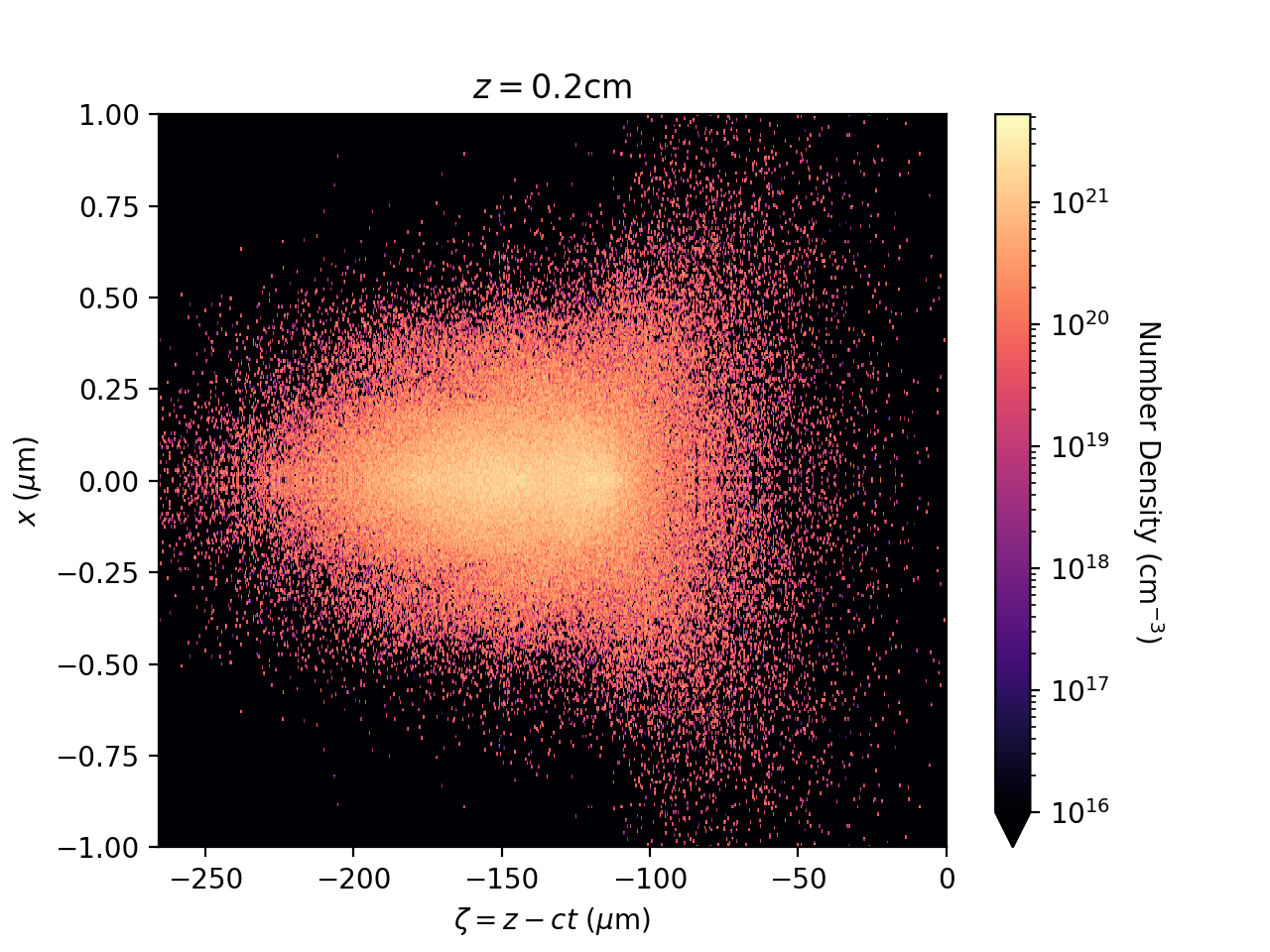}
    \includegraphics[width=0.25\textwidth]{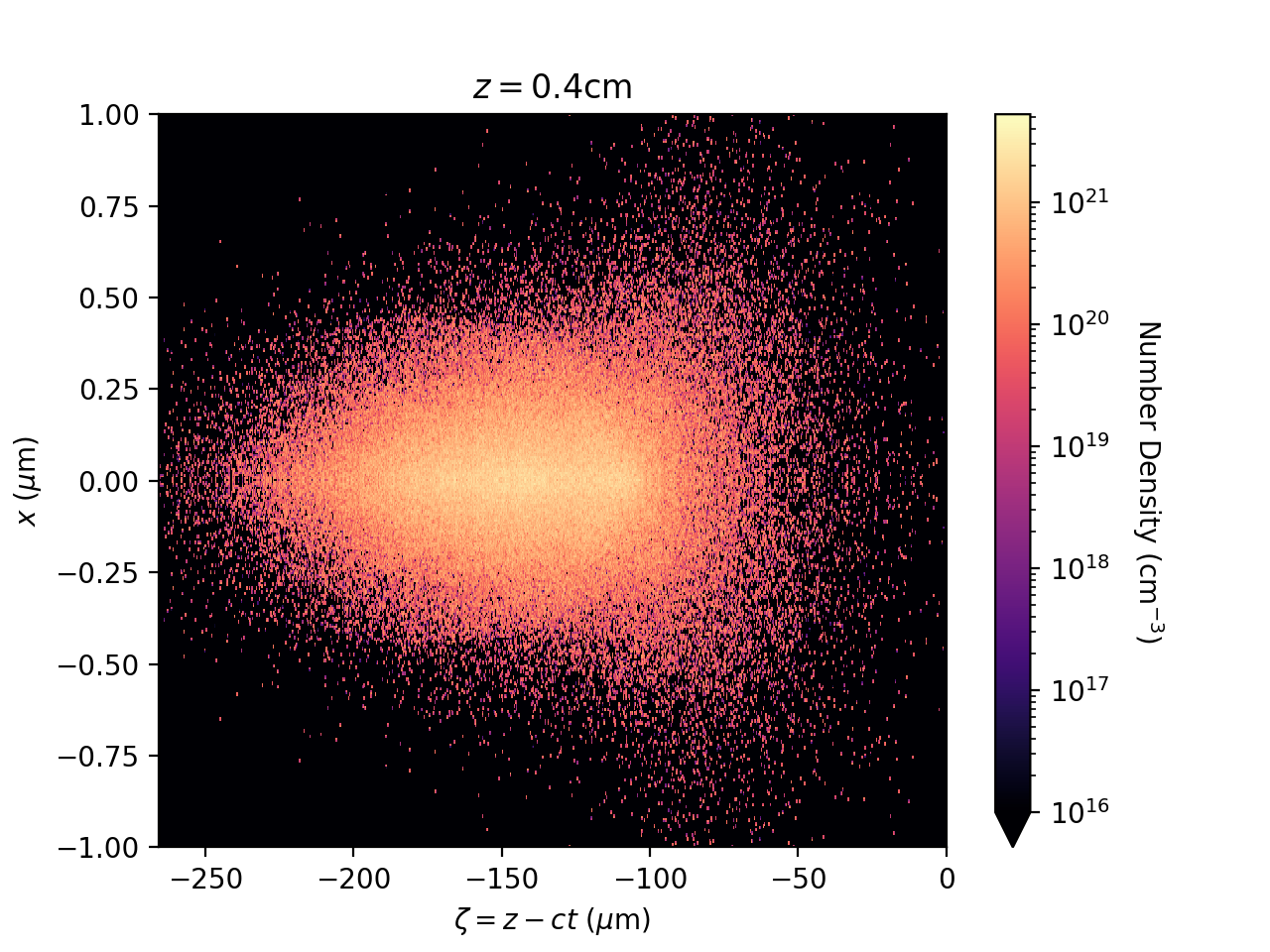}
    \includegraphics[width=0.25\textwidth]{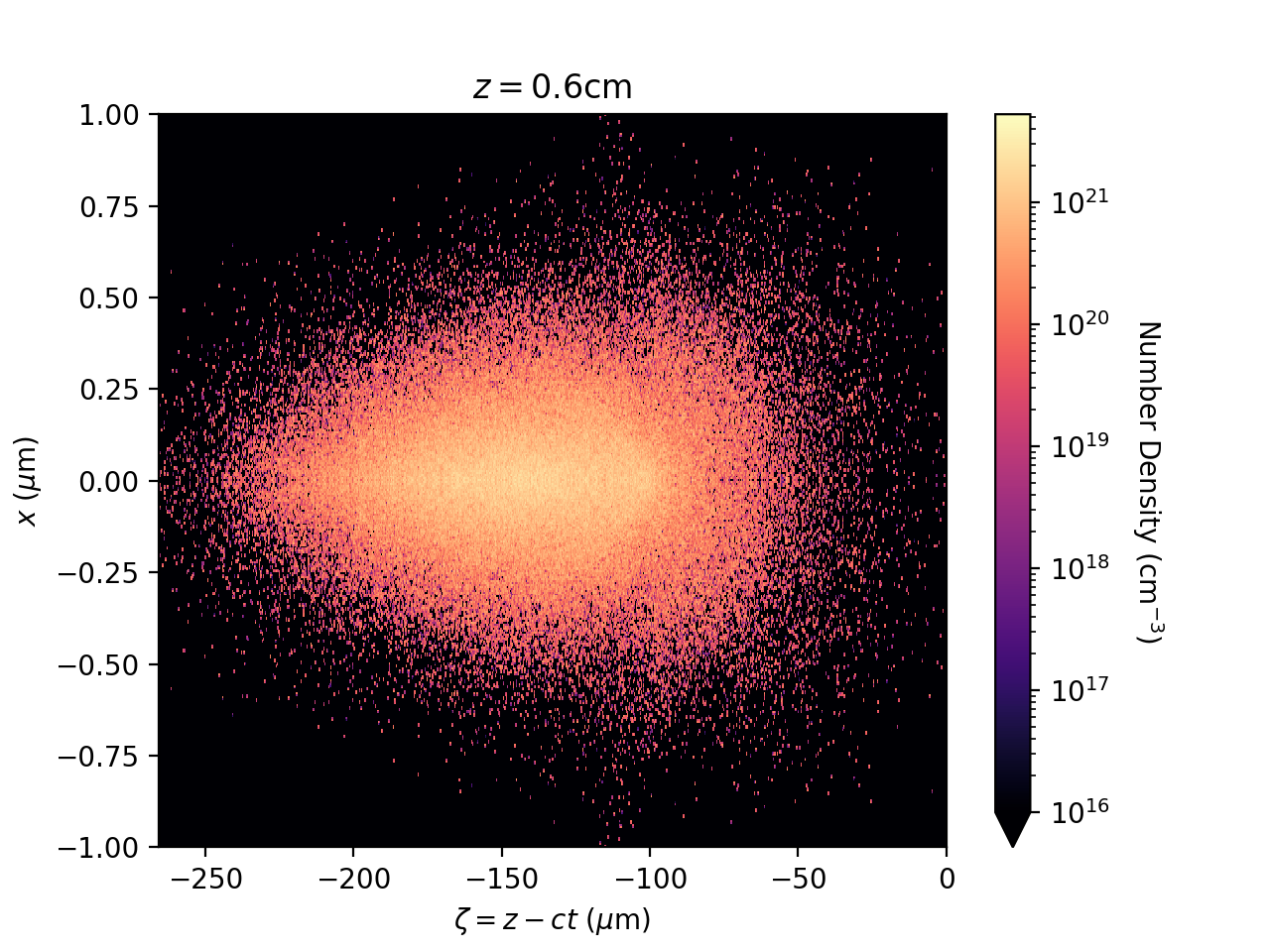}
    \centering
    \caption{Beam electron number density slice at $z = \SI{0}{mm}$ \textbf{(Far Left)}, $z = \SI{2}{mm}$ \textbf{(Middle Left)}, $z = \SI{4}{mm}$ \textbf{(Middle Right)}, and $z = \SI{6}{mm}$ \textbf{(Far Right)}.}
    \label{fig:beam}
\end{figure}

A high resolution simulation of the beam-ion interaction was conducted with similar parameters to the previous case except: (a) the approximation described above is used; (b) the transverse resolution was much higher with a transverse cell size of $\SI{5.2}{nm}$; (c) a smaller beam spot size of $\SI{200}{nm}$ was used which is closer to the spot size matched to the nonlinear focusing field; (d) a longer plasma of length $\SI{1.8}{cm}$ was used. The plasma ion number density distribution is shown on the left of Fig.~\ref{fig:other}. Compared to the first simulation, the ion column is both much denser and much thinner. In fact, neither of the simulations properly resolve the ion column; The number density appears to diverge, although the focusing fields do converge as the resolution is increased. The evolution of the emittance and beam spot size are also shown in Fig.~\ref{fig:other} and demonstrate decaying oscillations similar to the first simulation. The evolution of the beam number density is shown in Fig.~\ref{fig:beam}. Unlike the beam evolution of the previous simulation shown in Fig.~\ref{fig:e314beam}, phase space mixing occurs less at the tail of the beam and more at the head. This is explained by the fact that the initial beam spot size for this simulation is closer to being matched to the nonlinear focusing force and so little mixing occurs at the tail of the beam, but it is overfocused for the linear focusing force at the head of the beam which causes mixing there.

\begin{figure}[h]
    \centering
    \includegraphics[width=0.5\textwidth]{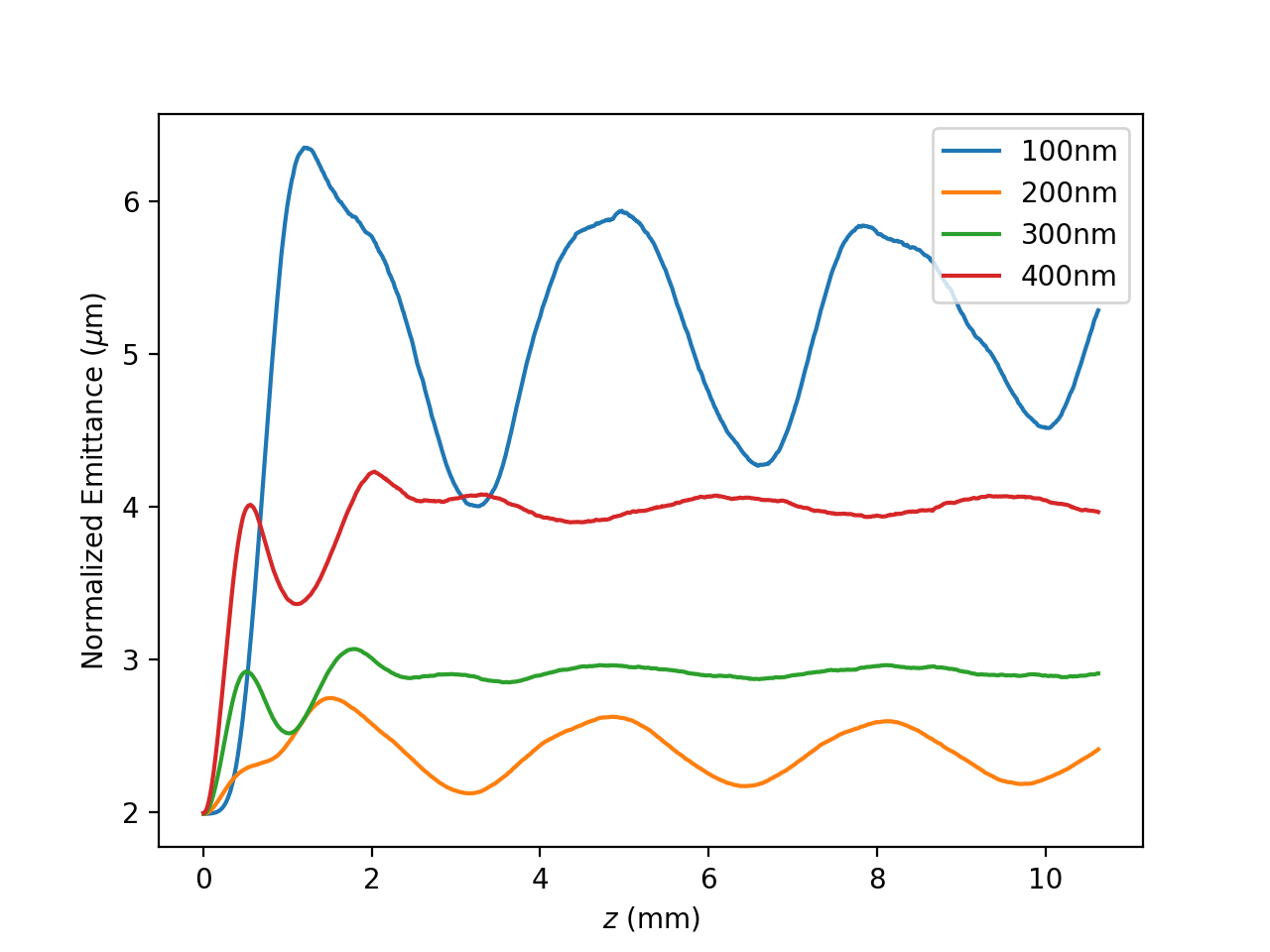}
    \includegraphics[width=0.5\textwidth]{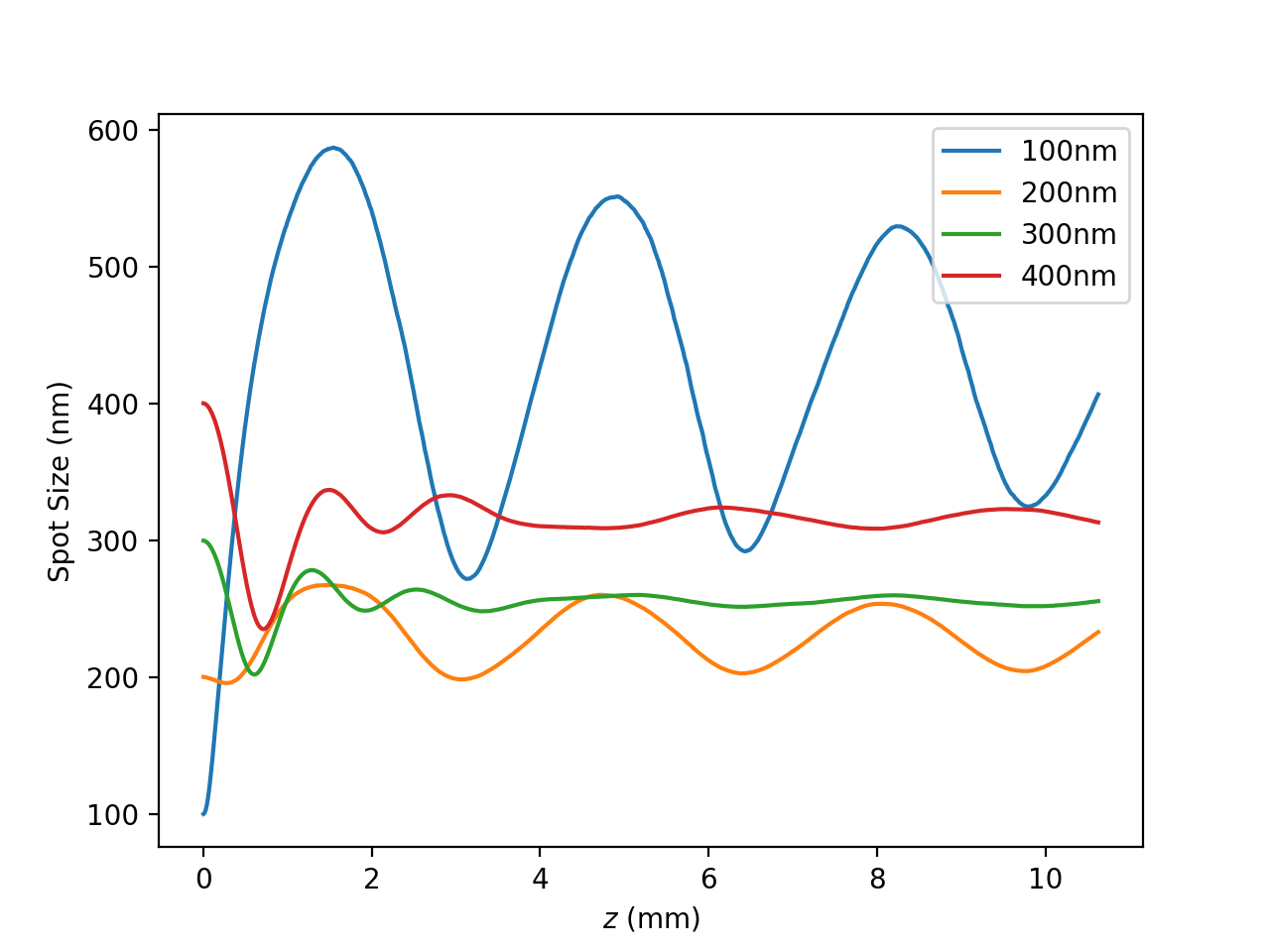}
    \centering
    \caption{\textbf{Left:} Beam emittance evolution over the length of the plasma for four different initial spot sizes. \textbf{Right:} Beam spot size evolution over the length of the plasma for four different initial spot sizes.}
    \label{fig:scan}
\end{figure}

Finally, we performed a scan over the initial spot size using parameters similar to this second simulation, albeit with a somewhat reduced resolution. The emittance and spot size evolution are shown in Fig.~\ref{fig:scan}. The four initial spot sizes are $\sigma = \SI{400}{nm}$ (significantly underfocused), $\sigma = \SI{300}{nm}$ (approximately linearly matched), $\sigma = \SI{200}{nm}$ (approximately nonlinearly matched), and $\sigma = \SI{100}{nm}$ (significantly overfocused). As would be expected in the absence of ion motion, the underfocused beam shows significantly larger emittance growth than the approximately linearly matched case. However, the beam with initial spot size $\sigma = \SI{200}{nm}$ which is smaller than the linearly matched case shows less emittance growth. This is the same result as \cite{an}. Finally, it is still possible to overfocus the beam which causes a huge increase in emittance.

\section{CONCLUSION}

In this paper we have discussed the physics of ion motion in plasma wakefield accelerators, outlined plans for the E-314 experiment at FACET-II, and presented PIC simulation results of plasma accelerators with ion motion relevant to the E-314 experiment. Understanding ion motion and the mitigation of emittance growth in plasma accelerators with ion motion is of paramount importance to the development of practical plasma based linear colliders. Experimental demonstration of ion column formation and emittance preservation via nonlinear matching is a crucial step towards this goal. E-314 will be the first experiment to be performed with these aims.

In the near future, commissioning of the FACET-II beamline and the first experiments will take place. Experimental hardware is currently being designed and developed for E-314 and other plasma acceleration experiments at FACET-II. Work is ongoing to develop an analytic model for the beam-ion equilibrium, and a particle tracking code to simulate emittance growth due to beam-ion scattering. Future work includes modeling betatron radiation for E-314 and understanding long term ion dynamics.

\section{ACKNOWLEDGMENTS}

This work was performed with support of the US Department of Energy, Division of High Energy Physics, under Contract No. DE-SC0009914, National Science Foundation under Grant No. PHY-1549132, DE-SC0009914 (UCLA), and the STFC Liverpool Centre for Doctoral Training on Data Intensive Science (LIV.DAT) under grant agreement ST/P006752/1. This work used computing resources provided by the STFC Scientific Computing Department’s SCARF cluster.

\nocite{*}
\bibliographystyle{aac}%
\bibliography{aac2020_latex}%

\end{document}